\documentclass[11pt,preprint]{aastex}

\def\plotfiddle#1#2#3#4#5#6#7{\centering \leavevmode
\vbox to#2{\rule{0pt}{#2}}
\includegraphics{#1}}

\slugcomment{To appear in the Astrophysical Journal, Letters}

\begin{document}
\title{Observations of DG Tauri with the Keck Interferometer}

\author {
M.~Colavita\altaffilmark{1}, 
R.~Akeson\altaffilmark{2}, 
P.~Wizinowich\altaffilmark{3},
M.~Shao\altaffilmark{1},
S.~Acton\altaffilmark{3},
J.~Beletic\altaffilmark{3},
J.~Bell\altaffilmark{3},
J.~Berlin\altaffilmark{1},
A.~Boden\altaffilmark{2},
A.~Booth\altaffilmark{1},
R.~Boutell\altaffilmark{3},
F.~Chaffee\altaffilmark{3},
D.~Chan\altaffilmark{3},
J.~Chock\altaffilmark{3},
R.~Cohen\altaffilmark{3},
S.~Crawford\altaffilmark{1},
M.~Creech-Eakman\altaffilmark{1},
G.~Eychaner\altaffilmark{1},
C.~Felizardo\altaffilmark{2},
J.~Gathright\altaffilmark{3},
G.~Hardy\altaffilmark{1},
H.~Henderson\altaffilmark{2},
J.~Herstein\altaffilmark{2},
M.~Hess\altaffilmark{3},
E.~Hovland\altaffilmark{1},
M.~Hrynevych\altaffilmark{3},
R.~Johnson\altaffilmark{1},
J.~Kelley\altaffilmark{1},
R.~Kendrick\altaffilmark{3},
C.~Koresko\altaffilmark{2},
P.~Kurpis\altaffilmark{3},
D.~Le Mignant\altaffilmark{3},
H.~Lewis\altaffilmark{3},
E.~Ligon\altaffilmark{1},
W.~Lupton\altaffilmark{3},
D.~McBride\altaffilmark{3},
B.~Mennesson\altaffilmark{1},
R.~Millan-Gabet\altaffilmark{2},
J.~Monnier\altaffilmark{4},
J.~Moore\altaffilmark{1},
C.~Nance\altaffilmark{3},
C.~Neyman\altaffilmark{3},
A.~Niessner\altaffilmark{1}
D.~Palmer\altaffilmark{1},
L.~Reder\altaffilmark{1},
A.~Rudeen\altaffilmark{3},
T.~Saloga\altaffilmark{3},
A.~Sargent\altaffilmark{2},
E.~Serabyn\altaffilmark{1},
R.~Smythe\altaffilmark{1},
P.~Stomski\altaffilmark{3},
K.~Summers\altaffilmark{3},
M.~Swain\altaffilmark{1},
P.~Swanson\altaffilmark{1},
R.~Thompson\altaffilmark{1},
K.~Tsubota\altaffilmark{3},
A.~Tumminello\altaffilmark{1},
G.~van Belle\altaffilmark{2},
G.~Vasisht\altaffilmark{1},
J.~Vause\altaffilmark{3},
J.~Walker\altaffilmark{3},
K.~Wallace\altaffilmark{1},
U.~Wehmeier\altaffilmark{3}
}
\altaffiltext{1}
{Jet Propulsion Laboratory, 4800 Oak Grove Dr., Pasadena, CA 91109; mcolavit@s383.jpl.nasa.gov}
\altaffiltext{2}
{Michelson Science Center, 770 S. Wilson Ave., Pasadena, CA 91125; rla@ipac.caltech.edu}
\altaffiltext{3}{W.M.Keck Observatory, California Association for Research in Astronomy, 65-1120 Mamalahoa Hwy., Kamuela, HI 96743; peterw@keck.hawaii.edu}
\altaffiltext{4}{Univ. of Michigan, 941 Dennison Bldg, Ann Arbor, MI, 48109}


\begin{abstract}
  
We present the first science results from the Keck Interferometer, a
direct-detection infrared interferometer utilizing the two 10-meter
Keck telescopes.  The instrument and system components are briefly
described.  We then present observations of the T~Tauri object DG~Tau,
which is resolved by the interferometer.  The resolved component has a
radius of 0.12 to 0.24 AU, depending on the assumed stellar and
extended component fluxes and the model geometry used.  Possible
origins and implications of the resolved emission are discussed.

\end{abstract}

\section{Introduction}

The sensitivity provided by using large aperture telescopes in
optical/infrared interferometers will greatly expand the range of
astronomical sources which can be studied on milliarcsecond (mas)
scales.  One such facility is the Keck Interferometer (KI), and we
present its first science results: observations of the young stellar
object DG~Tau.

The Keck Interferometer\footnote[5]{Additional information at
http://planetquest.jpl.nasa.gov/} is a NASA-funded joint development
among the Jet Propulsion Laboratory (JPL), the California Association
for Research in Astronomy, and the Michelson Science Center at the
California Institute of Technology\footnote[6]{Additional information at
http://msc.caltech.edu/},
to interferometrically combine the two 10 meter Keck telescopes for high
sensitivity near-infrared(IR) visibility amplitude measurements, mid-IR 
nulling interferometry at $10 - 12~\mu$m, and differential-phase
interferometry at $1.6 - 5$~$\mu$m. First fringes were obtained in
2001 March using the two Kecks with their adaptive optics (AO)
systems, with subsequent activities directed toward improving
visibility-mode performance and commencing shared-risk science.

\section{Instrument Description}

Instrument details are given by \citet{col02}; in brief, KI uses
pupil-plane combination between the two 10-m Keck telescopes,
providing an 85-m baseline oriented at 38$^{\circ}$ (E of N).  Both
telescopes have adaptive optics systems \citep{wiz02}, which are used
for all interferometer observations.  For the measurements described
here, a single beam from each telescope is routed through the
telescope coude train to a beam combining laboratory.  Optical path
delay is implemented in two stages, with long delay lines (LDLs),
which are stationary during an observation, and fast delay lines
(FDLs), which track sidereal motion.  The available delay on the sky
for a given LDL position is $\pm$15~m.  The FDLs incorporate
measurements from local laser metrology, an end-to-end laser metrology
system, and telescope accelerometers for fast servo control and to
stabilize the optical path.  The KI control system is described by
\citet{boo02}.

Angle tracking to stabilize the images from the individual telescopes
is implemented using a HAWAII near-IR array camera system which
controls fast tip/tilt mirrors in the beam combining lab.  For the
data presented here, the angle tracker operated at $J$(1.2 $\mu$m)
with a 100~Hz frame rate.  Fringe tracking and science measurements
use a second HAWAII near-IR array camera system which controls
the FDLs \citep{vas02}.  The camera is fed using single-mode fluoride
optical fibers from a free-space beam-combination breadboard, and
provides a white-light channel from one beamsplitter output and a
low-resolution spectrometer channel from the complementary output.
The (single-mode) field-of-view on the sky is $\sim$50 mas (FWHM) at
2.2~$\mu$m.  The fringe tracker implements a 4-bin synchronous fringe
demodulation algorithm similar to that used at the Palomar Testbed
Interferometer (PTI) \citep{col99b}.  For the data presented
here, the system operated at $K^{\prime}$, with 4-pixel dispersion on
the spectrometer side, and used a 500~Hz frame rate.

\section{Results}

\subsection{DG Tau}
\label{dgtau}

T Tauri stars are pre-main sequence objects with stellar masses less
than $\sim$2 M$_{\odot}$ and ages of 1 to 10 million years.  The
canonical model for these sources comprises a central star surrounded
by a circumstellar disk of size $\sim$100~AU, a collimated jet or
outflow, and perhaps some residual material in a more extended
envelope.  Due to the high resolution required (1~AU = 7
mas at 140 pc), {\it
direct} observations of the central region have been
limited to date.  Characterizing the inner disk properties
is important for understanding hydrodynamic disk
winds and the initial conditions of planet
formation.  Given the spatial resolution of current facilities,
long-baseline infrared interferometry provides an ideal method for
observing these inner regions, which are traced by material thermally
emitting in the near-IR.

Previous interferometric observations have resolved several types of
young stellar objects (YSO): FU Ori \citep{mal98}, T Tauri
\citep{ake00}, and Herbig AeBe sources \citep{mil99,mil01}.  For the T
Tauri and Herbig objects, the measured visibilities are generally not
consistent with a geometrically flat disk with an inner radius less
than a few stellar radii as predicted by 
spectral energy distribution fitting.  \citet{tut01} and \citet{nat01}
proposed that the inner radius is located at the destruction radius for
directly heated dust, which can resolve this discrepancy.  However,
due to the sensitivity limits of the facilities used, the YSOs
observed to date, particularly the T~Tauris, are the brightest
infrared examples and may not be representative.

Here we present observations of DG~Tau (d=140~pc), a T Tauri star with a more
typical luminosity, which has a well-studied jet (see
e.g. \citet{lav00,bac02} and references therein) and a
circumstellar disk observed in both dust continuum and molecular gas
\citep{kit96a,kit96b,tes02}.  The infrared spectral index for DG~Tau
is roughly flat, placing it in the class II evolutionary category 
\citep{lad87}.  DG Tau has a $K$
magnitude of 6.98 $\pm$0.013 (2MASS, 1997 November 30) and an average
$V$ magnitude of 12.43 \citep{ken95}.  The optical jet has a position
angle (PA) of 226\arcdeg\ and a derived inclination of 38\arcdeg\ from
the line of sight \citep{bac02}.  On larger scales, the disk
inclination has been estimated as 70\arcdeg\ at $\sim$100~AU and at
40\arcdeg\ at $\sim$2800~AU \citep{kit96a,kit96b}.

\subsection{Observations and data reduction}
\label{data}

DG~Tau was observed with the Keck Interferometer on 2002 October 23
and 2003 February 13 UT.  The data presented here are from the
white-light channel ($\lambda_{\rm center}$=2.18~$\mu$m and $\Delta
\lambda \sim$~0.3$\mu$m).  Observations consisted of a series of
interleaved integrations on the source and several calibrators.  Each
integration includes 120 seconds of fringe data followed by
measurements of the background flux, the individual fluxes in the two
arms of the interferometer, and the foreground flux.  The data 
presented are the visibility amplitude squared, an unbiased
quantity, normalized such that an unresolved object has V$^2=1.0$.

The system visibility (V$_{sys}^2$), i.e., the instrumental response to a point
source, is measured with respect to calibrator stars HD 282230, HD
283668, and HD 29050.  The calibrators are located 5.3\arcdeg,
1.7\arcdeg\ and 3.6\arcdeg\ from DG~Tau, respectively, and have $K$
magnitudes of 7.3, 7.0 and 7.1, well matched to DG~Tau's
$K$ magnitude.  The calibrator angular sizes were derived by
fitting photometry from SIMBAD and 2MASS.  All calibrators have
angular diameters $<$ 0.3 mas, and are unresolved by the
interferometer.  Source and calibrator data were corrected for
detection biases as described by \citet{col99} and averaged into
5 second blocks. V$_{sys}^2$ is calculated for
each DG Tau data point using an average of the calibrator measurements
weighted by the internal scatter in the calibrator data and the
temporal and angular proximity to the target data point
\citep{bod98}. In addition, the V$_{sys}^2$ error includes the
uncertainty in the calibrator size, assumed to be 0.1 mas.  The
calibrated data points for the target source are the average of the 5
sec blocks in each integration, with an uncertainty given by the
quadrature of the internal scatter and the uncertainty in V$_{sys}^2$.
The average V$^2$ for DG Tau is 0.37 with a
measurement uncertainty of 0.02.

Several checks were performed to verify the measured visibility.  The
calibrator $K$ magnitudes are within 0.3 magnitudes of DG~Tau,
minimizing any errors due to flux dependencies in the fringe tracker.
For comparison to the white light data presented, a synthetic
wide-band measurement was constructed by summing the four spectrometer
channels.  These data are within 1$\sigma$ of the white-light data.
In addition to measurement errors, there may be systematic errors in
V$^2$.  These have been estimated using observations of binary
systems with known orbital parameters.  These binary systems had V$^2$
of 0.2 to 0.4, similar to the DG~Tau visibility, and an rms from the
predicted visibility of 0.05.  Therefore, as a conservative estimate
of the KI performance, we assign a systematic error of 0.05 to the
DG~Tau data and quadratically sum this error with the measurement
uncertainties (Fig. 1).  The systematic error dominates the
measurement error, resulting in an average V$^2$ for DG~Tau of
0.37$\pm$0.05 at a projected baseline length of 84.7 meters.

\begin{figure}[h!]
\plotfiddle{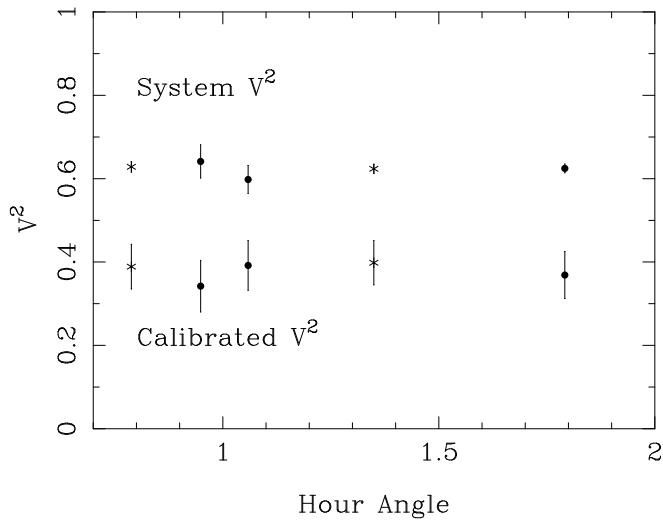}{2.5in}{270}{70}{70}{-270}{330}
\caption{The calibrated DG~Tau visibilities (bottom points) and system
visibilities (top points).  Data from 2002 October are labeled with
stars and from 2003 February with circles.  The projected baseline
ranges from 84.2 to 84.9 meters at PA 32$^{\circ}$ to 25$^{\circ}$ (E
of N).  The DG~Tau error bars include both the measurement and
systematic components.
\label{fig:data}}
\end{figure}

\subsection{Models of source structure}
\label{models}

As seen in Fig.~1, DG~Tau is clearly resolved by KI.  We have used
simple geometric models of the source brightness to estimate the size
scale of the infrared emission region.  Several components may be
contributing to the measured visibility: the stellar photosphere,
which is unresolved in our observations; emission from compact
material, which is less than the 5 mas fringe spacing and is partially
resolved; and extended emission, which is completely resolved and
contributes incoherently.

Before the size of the compact component can be determined, the
stellar and incoherent contributions must be quantified.  In the
spectrum of DG~Tau, the line features are veiled by continuum
emission, making the spectral type difficult to determine, with a
range in the literature of K7 to M0 \citep{ken95}.  We have used two
methods to calculate the stellar contribution to the $K$ flux: 1)
using the effective temperature (T$_{\rm eff}$=3890~K) and extinction
(A$_{\rm V}$ = 1.6) from \citet{bec90} and fitting the de-reddened
optical photometry, which gives a stellar/total flux ratio of 0.21; and
2) simply adopting the flux ratio of 0.41 derived by \citet{joh01}
using infrared spectroscopy.  To estimate any extended component, we
use the lunar occultation observations of \citet{lei91} and
\citet{che92}.  These measurements are sensitive to emission on scales
from 5 mas to 1\arcsec.  \citet{lei91} fit the extended component at $K$
with a 45 mas (FWHM) Gaussian with a flux ratio
(extended/total) of 0.23.  \citet{che92} derived similar results from
independent observations.  \citet{lei91} also used
speckle interferometry, which revealed a 0\farcs85 symmetric envelope.
As this larger envelope contributes only
3\% of the $K$ flux, we do not include it in our modeling.
Based on the flux and size of the 45 mas emission (hereafter,
the extended component), \citet{lei91} suggest that this emission is
likely due to scattered light rather than thermal emission.  

We chose two simple geometries to estimate the size of the compact
component.  Given the observed non-variation of the visibilities with
baseline orientation within the limited hour angle range of our
observations, we cannot constrain any asymmetry in the source, and
therefore consider only symmetric models -- a uniform disk and a
uniform brightness ring (Table 1).  The ring width is set for a given inner
diameter by requiring that the ring flux for a typical dust
destruction temperature (1500~K) match the observed $K$ excess for the
compact component. 
Both the stellar (unresolved) and extended (completely resolved)
components are included in the modeling.

\begin{table}[h!]
\begin{center}
\begin{tabular}{lll} \tableline
F$_{\rm star}$/F$_{\rm total}$ & 0.41 & 0.21 \\ \tableline
UD model \\
\quad radius (mas) & 1.71$\pm$ 0.26 & 1.33$\pm$0.19 \\
\quad radius (AU) & 0.24$\pm$0.036 & 0.19$\pm$0.027 \\
Ring model \\
\quad inner radius (mas) & 1.16$\pm$0.17 & 0.89$\pm$0.14 \\
\quad inner  radius (AU) & 0.16$\pm$0.024 & 0.12$\pm$0.020 \\
\quad width (AU) & 0.006 & 0.012 \\ \tableline
\end{tabular}
\caption{The fit radii and uncertainties (including measurement
and systematic errors) for the uniform disk (UD) and ring models for
two values of the stellar fractional flux (\S \ref{models}).  All
models include an incoherent component with a fractional flux of 0.23.
The ring model width is set as described in the text for a blackbody
temperature of 1500~K.   }
\end{center}
\end{table}

\section{Discussion}

We have estimated the radius of the compact emission region of DG~Tau to
be 0.12 to 0.24 AU given our estimates for the relative flux of the
stellar and extended components.  Here we discuss the hypothesis that
this resolved component represents thermal emission from the inner
regions of the circumstellar disk.  Using a standard T~Tauri disk
model with a temperature power-law $T \propto r^{-q}$, T(1 AU) =
150~K and q=0.5 (see e.g. \citet{bec90}), the 2 $\mu$m emission arises
from material located within 5 mas of the star.

The radii in Table 1 are similar those derived for the disks around the T~Tauri
stars T~Tau~N and SU~Aur by \citet{ake00}, and all are larger than
predicted by models based on spectral energy distribution fitting.  We
use the ring morphology as a simple representation of models where
the inner ring radius is set by the dust destruction radius
\citep{dul01,mon02}, where $R_{\rm dust}=1/2(L_{\star}/4 \pi \sigma
T_{\rm dust}^4)^{1/2}$.  Figure 2 shows the radius corresponding to a
dust destruction temperature (T$_{\rm dust}$) of 1500~K plotted
against the fit ring radius for DG~Tau.  In addition, four T~Tauri
sources observed at PTI are also plotted \citep{ake02}.  For DG~Tau
the observed radius is {\it larger} than $R_{\rm dust}$ for 
T$_{\rm dust}$=1500~K, while for the PTI sources the observed radius is
similar or smaller.  Although there are too few sources to
generalize, DG~Tau is the faintest infrared source of the group.

\begin{figure}[h!]
\plotfiddle{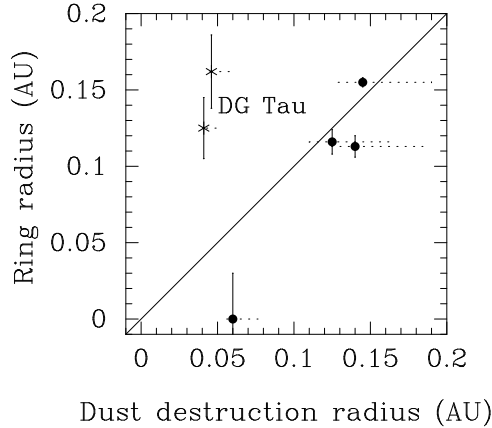}{2.1in}{270}{80}{80}{-170}{440}
\caption{A comparison of the dust destruction radius (for T$_{\rm
dust}$=1500~K) and the best fit ring model (inner radius) for
DG~Tau (stars) and four other T Tauri sources observed with PTI
(circles).  The solid line illustrates the location where these radii
are equal.  The two DG~Tau points correspond to the two stellar models
used.  The horizontal dotted lines represent a
typical range of T$_{\rm dust}$ (1200 to 1800~K).  The
y-axis error bars do not include the uncertainty in the fractional flux
from the unresolved and extended components.
\label{fig:ring}}
\end{figure}

We now briefly mention several issues and caveats in our determination of
the size of the $K$ emission region for DG~Tau.

{\bf Source variability:} DG~Tau is known to vary at optical and
infrared wavelengths; \citet{skr96} measured the $K$ variability to be 0.8 mag.
If this variability changes the relative contributions of the stellar,
compact and extended components, then the derived size is
directly affected.  For example, if the scattered light emission at $K$ were
0.39 times the total flux (1.7 times higher than measured by
\citet{lei91}), the measured visibility would be accounted for by just
the stellar (unresolved) and extended components.  Although this level of
variability is seen in the total $K$ flux, the ratio of scattered to
stellar flux should be constant for a brightening of the star.

{\bf Extended flux contribution:} We have assumed that the 45 mas
extended component is circularly symmetric and completely coupled into
the fringe tracker.  The exact coupling of this emission into the
fringe tracker field-of-view is dependent on the extended component
shape and on the AO and system performance.  The lunar occultation
observations of \citet{lei91} and \citet{che92}, taken at PAs of
90\arcdeg, 233\arcdeg\ and 256\arcdeg, resulted in similar size
determinations, suggesting this component is roughly symmetric.  Any
reduction in the extended component coupling would result in a {\it
larger} size for the partially resolved component.  If there were no
extended component, we would obtain a ring radii of 0.17 to 0.21~AU
(for the two stellar models).  Scattered light may be affecting the
stellar properties derived by \citet{bec90} and \citet{joh01}.  A
smaller stellar (unresolved) component would result in a {\it smaller}
ring radius.

{\bf Inclination:} For an inclined disk, a less
simplistic representation is necessary as the \citet{nat01} models are
vertically extended at the inner edge.  However, in their T~Tauri
models, the height is much less than the radius at the inner edge, and
therefore the measured size will underestimate the physical size.  If
the inner disk is orthogonal to the jet (which implies a disk PA of
136\arcdeg ), the actual size will be only a few percent larger than
our measured size given the measurement PA.  However, if the DG~Tau
disk is near edge-on, scattered light in the polar regions
could be more significant, increasing the measured size.

{\bf Dust destruction radius:} The dust destruction radius depends on
the assumed stellar luminosity and on T$_{\rm dust}$.  For the dust
destruction radius to match the measured ring radius at T$_{\rm dust}$=1500~K,
a luminosity of 3 to 5 L$_{\odot}$ would be required, while
the values for the assumed stellar models are 1.7 L$_{\odot}$
\citep{bec90} and 1.4 L$_{\odot}$ \citep{joh01}.  Similarly, for the
luminosity values used, a dust destruction temperature of T$_{\rm
dust}$= 1100 to 1300~K would be required to match the measured ring
radius.  As discussed by \citet{mon02}, small grains have a larger
dust destruction radius.

{\bf Multiplicity:} An additional point source within the KI
field-of-view (50 mas) could account for the measured visibility.
\citet{lei91} observed DG~Tau with speckle interferometry and found no
companion source.  Although we see no evidence for visibility
variability, as would occur for a coherently contributing source, the
presence of a companion can not be completely ruled out from the
present KI data.

If our model of the extent and brightness of the DG~Tau infrared
emission components is correct, the difference between the estimated
ring radius and the radius at which T$_{\rm dust}$=1500~K is
significant.  The V$^2$ for DG~Tau is 0.37$\pm$0.05,
while the
stellar properties of DG~Tau predict a V$^2$ of 0.54 to 0.57
(including the extended component) for T$_{\rm dust}$=1500~K.
Although there may be optically thin gas or dust within the derived
ring radius, a large gap between the star and the optically thick
inner edge of the disk has implications for some hydrodynamic wind
models and planet migration theories. 

We have demonstrated the basic capabilities of the KI in V$^2$ mode,
using the interferometric combination of two AO-equipped 10 meter
class telescopes. We present the first KI science result: the
detection of clearly resolved near-IR emission from DG~Tau, the
faintest T Tauri object measured to date with an optical/IR
interferometer.   
These observations provide
constraints on theories of circumstellar disks and demonstrate the
potential of large aperture optical/IR interferometry in providing
information on milliarcsecond scales not only for young stellar
objects, but for many areas of stellar and extra-galactic
astrophysics.

\acknowledgments

KI is funded by the National Aeronautics and Space
Administration(NASA).  Part of this work was performed at the Jet
Propulsion Laboratory, California Institute of Technology(Caltech) and at
the Michelson Science Center(MSC), under contract with NASA.
Observations were obtained at the W.M. Keck Observatory,
operated as a scientific partnership among Caltech,
the University of California and NASA.  The Observatory
was made possible by the generous financial support of the W.M. Keck
Foundation.  The authors wish to recognize and acknowledge the very
significant cultural role and reverence that the summit of Mauna Kea
has always had within the indigenous Hawaiian community.  We are most
fortunate to have the opportunity to conduct observations from this
mountain.  This work has used software from the MSC; the SIMBAD
database, operated at CDS, Strasbourg, France; and the NASA/IPAC
Infrared Science Archive, operated by JPL under contract with NASA.


\end{document}